\documentclass{iau}

\usepackage{amsmath}
\usepackage{graphicx}
\usepackage{multirow}
\usepackage{subcaption}
\usepackage{caption}

\begin{document}
\lefttitle{Cambridge Author}
\righttitle{SERVIMON: AI-Driven Predictive Maintenance and Real-Time Monitoring for Astronomical Observatories}

\jnlPage{x}{y}
\jnlDoiYr{2025}
\doival{10.1017/xxxxx}

\aopheadtitle{IAUS 397: Exploring the Universe with Artificial Intelligence (UniversAI)}
\editors{C. Sterken,  J. Hearnshaw \&  D. Valls-Gabaud, eds.}

\title{SERVIMON: AI-Driven Predictive Maintenance and Real-Time Monitoring for Astronomical Observatories}

\author{Emilio Mastriani$^1$, Alessandro Costa$^1$, Federico Incardona $^1$, Kevin Munari$^1$, Sebastiano Spinello$^1$}
\affiliation{$^1$INAF, Osservatorio Astrofisico di Catania Via S Sofia 78, I-95123 Catania, ITALY\\ email: {\tt emilio.mastriani@inaf.it}}

\begin{abstract}
Objective: ServiMon is designed to offer a scalable and intelligent pipeline for data collection and auditing to monitor distributed astronomical systems such as the ASTRI Mini-Array. The system enhances quality control, predictive maintenance, and real-time anomaly detection for telescope operations.
Methods: ServiMon integrates cloud-native technologies—including Prometheus, Grafana, Cassandra, Kafka, and InfluxDB—for telemetry collection and processing. It employs machine learning algorithms, notably Isolation Forest, to detect anomalies in Cassandra performance metrics. Key indicators such as read/write latency, throughput, and memory usage are continuously monitored, stored as time-series data, and preprocessed for feature engineering. Anomalies detected by the model are logged in InfluxDB v2 and accessed via Flux for real-time monitoring and visualization.
Results: AI-based anomaly detection increases system resilience by identifying performance degradation at an early stage, minimizing downtime, and optimizing telescope operations. Additionally, ServiMon supports astrostatistical analysis by correlating telemetry with observational data, thus enhancing scientific data quality. AI-generated alerts also improve real-time monitoring, enabling proactive system management.
Conclusion: ServiMon’s scalable framework proves effective for predictive maintenance and real-time monitoring of astronomical infrastructures. By leveraging cloud and edge computing, it is adaptable to future large-scale experiments, optimizing both performance and cost. The combination of machine learning and big data analytics makes ServiMon a robust and flexible solution for modern and next-generation observational astronomy.
\end{abstract}


\maketitle

\section{Introduction}

The ASTRI Mini-Array  \cite{Giuliani} represents a significant step forward in ground-based gamma-ray astronomy, consisting of multiple small-sized telescopes operating in a distributed configuration. Designed as a precursor to the larger Cherenkov Telescope Array \cite{Actis2011}, the ASTRI Mini-Array serves as both a technological pathfinder and a scientific instrument in its own right. Its architecture enables high-throughput, multi-telescope observations, but also introduces challenges in maintaining system reliability, ensuring data integrity, and supporting uninterrupted operations across geographically dispersed components. Given the volume and velocity of telemetry generated by such distributed systems, robust monitoring, logging, and fault-detection mechanisms are essential for operational continuity. These tools must handle complex telemetry streams, enable real-time analytics, and support early detection of anomalies to reduce downtime and optimize performance. To address these needs, we introduced  ServiMon \cite{Munari2025}, a scalable, Docker-based \cite{Docker} data collection and monitoring pipeline specifically designed for complex environments like the ASTRI Mini-Array. ServiMon integrates cloud-native technologies—including Prometheus \cite{RabensteinVolz}, Grafana \cite{Grafana}, Loki \cite{GrafanaLoki}, Promtail \cite{Promtail}, Cassandra \cite{ApacheCassandra}, and Kafka  \cite{ApacheKafka}—to deliver real-time system monitoring, interactive visualization, and intelligent fault detection. This paper presents ServiMon’s architecture and implementation, showing how it can enhance system reliability and scalability while laying the foundation for predictive maintenance in next-generation astronomical infrastructures.

\section{System Architecture and Core Technologies}

ServiMon is built upon three foundational pillars: a Cloud-Native Stack, a Machine Learning Core, and Real-Time Processing capabilities. From the cloud-native perspective, ServiMon integrates technologies such as Prometheus, Grafana, Cassandra, Kafka, and InfluxDB \cite{InfluxDB} to enable comprehensive telemetry collection and scalable data processing across distributed astronomical infrastructures. The Machine Learning Core leverages the Isolation Forest \cite{IsolationForest} algorithm to detect anomalies in Cassandra performance metrics, continuously monitoring key indicators such as read/write latency, throughput, and memory usage, all captured as time-series data. In terms of real-time processing, performance metrics are preprocessed for feature engineering, with detected anomalies stored in InfluxDB v2 and accessed via Flux \cite{Flux} queries to support immediate visualization and system-level responses. The sequence diagram in Figure 1 reports the interaction among the three blocks.

\begin{figure}
    \centering
    \includegraphics[width=1\linewidth]{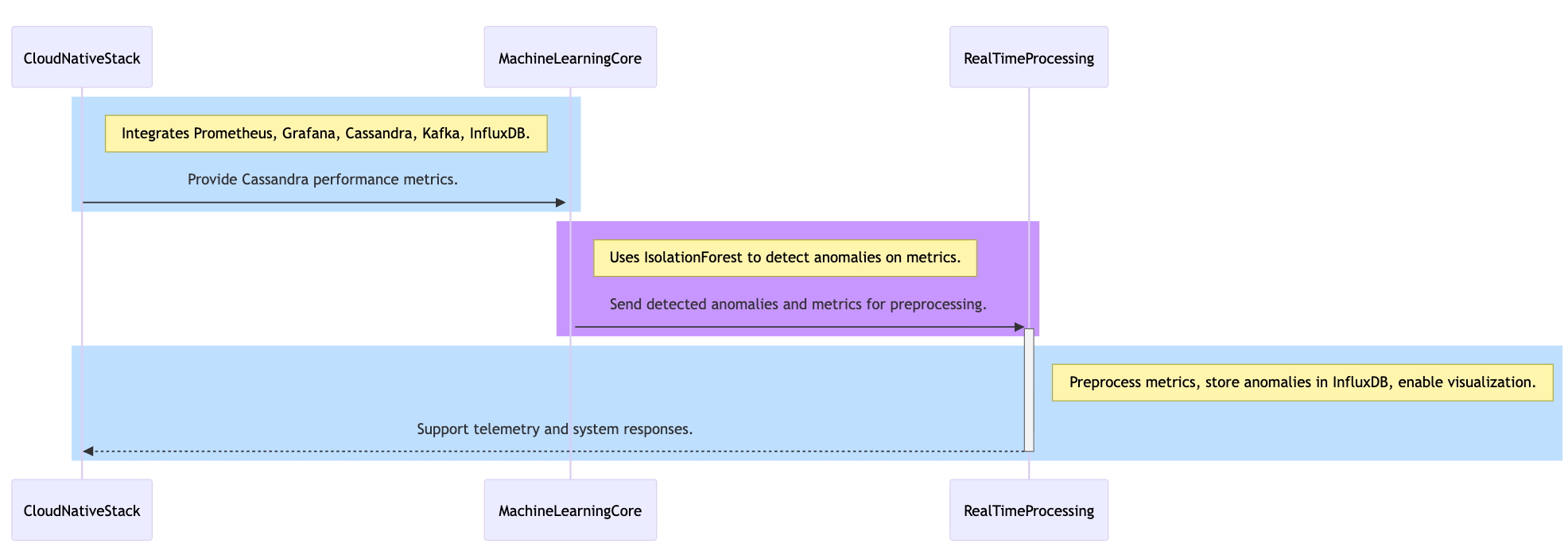}
    \caption{Three blocks interaction}
    \label{fig:placeholder}
\end{figure}

\section{Data Flow and System Integration}
The ServiMon architecture establishes a streamlined data flow for efficient metric collection, storage, and visualization. The process begins with the \textbf{metric exposure} phase, where the storage container makes Prometheus-style metrics available at the endpoint \verb|1235/metrics|, enabling continuous monitoring access. In the \textbf{data collection} stage, a Telegraf\cite{Telegraf} container retrieves these metrics over HTTP using the \verb|inputs.prometheus| plugin, ensuring seamless integration with the monitoring pipeline. During \textbf{storage processing}, Telegraf forwards the collected metrics to the \verb|cassandra_metrics| bucket in InfluxDB 2.x via the \verb|outputs.influxdb_v2| plugin. Finally, in the \textbf{visualization access} phase, InfluxDB stores the time-series data, which can be queried and visualized through Grafana dashboards, supporting real-time system monitoring and analysis. Figure 2 (a) shows the complete data flow. 

\begin{figure}[htbp]
    \centering
    \begin{subfigure}{\linewidth}
        \includegraphics[width=0.75\linewidth]{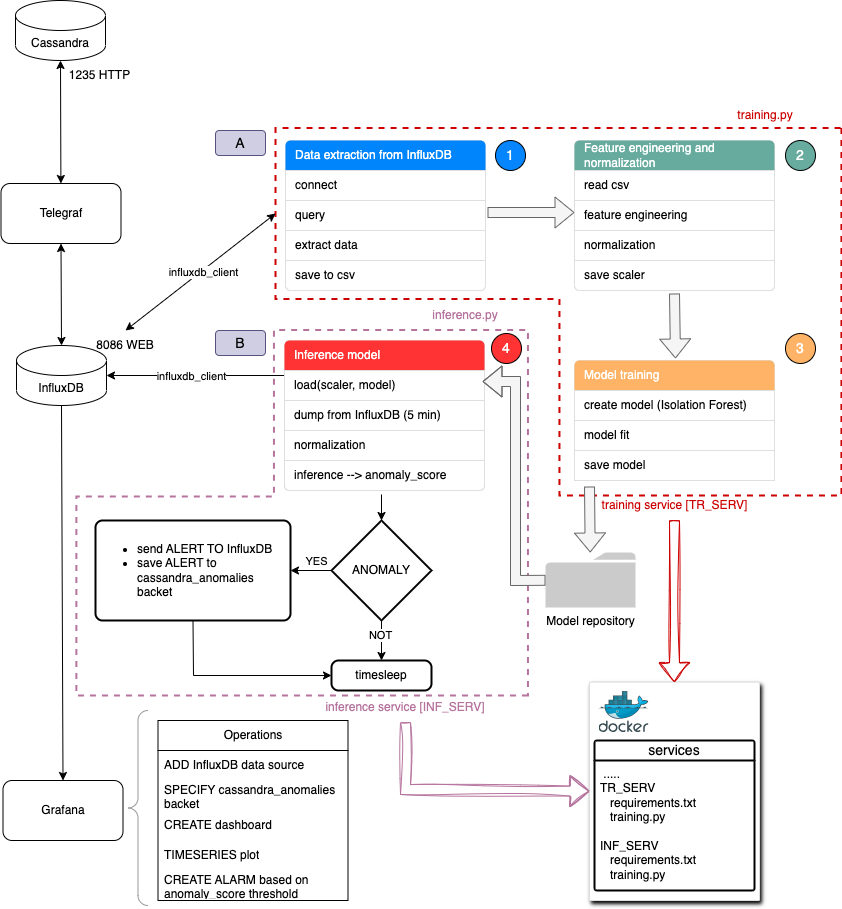}
        \caption{Data Flow and System Integration}
        \label{fig:prometheus-flow}
    \end{subfigure}

    \vspace{0.5cm} % Spazio verticale tra le due figure (puoi regolarlo)

    \begin{subfigure}{\linewidth}
        \includegraphics[width=\linewidth]{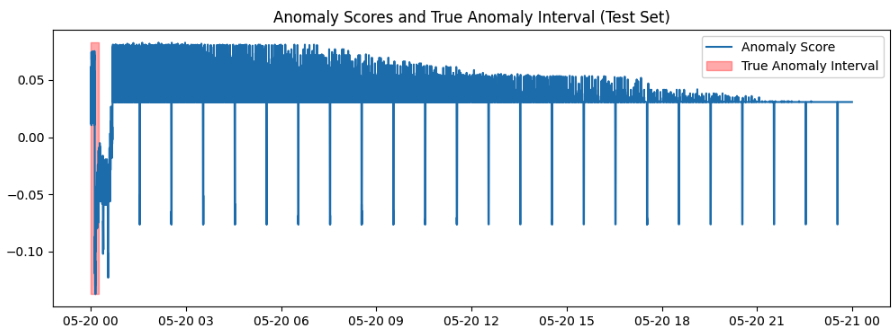}
        \caption{Anomaly detection on testing dataset}
        \label{fig:anomaly-detection}
    \end{subfigure}

    \caption{System overview and ML anomaly detection results}
    \label{fig:combined}
\end{figure}

\section{Machine Learning Model Implementation}
The implemented machine learning model is structured to support predictive maintenance for Cassandra by analyzing real-time telemetry data. Its architecture comprises two distinct modules: a \textbf{Training Module} and an \textbf{Inference Module} which operate independently. The Training Module periodically acquires historical telemetry data from InfluxDB, preprocesses it through a defined feature engineering pipeline (including scaling, feature selection, and NaN handling), and trains the model—typically an Isolation Forest—while optimizing hyperparameters. The resulting pipeline, including preprocessing and model, is saved in a portable format (e.g., `.pkl`). This process is initially manual but becomes automated for subsequent retraining cycles using updated data. In contrast, the Inference Module functions in an event-driven manner, typically executing hourly. It loads the most recent model version, queries real-time Cassandra and JVM metrics from Prometheus or InfluxDB, and applies the trained pipeline to detect anomalies. The inference results are then written back to InfluxDB along with their corresponding timestamps. This modular design ensures scalable, maintainable, and timely detection of system anomalies for proactive maintenance.

\section{Testing and Results}

Two distinct test phases were conducted to evaluate the system: one for model training and another for validating the inference and alert generation mechanism.

 With regards to the \textbf{training }phase, the baseline telemetry was generated using four \verb|opcuasimulatormon| container instances, simulating 998 monitoring points over a 24-hour period. To introduce variability, two stress sessions were injected using the \textit{home-made} \verb|cassandra-traffic| container, each lasting 15 minutes and simulating fault events at approximately 2\% frequency. The resulting dataset, exhibiting a typical imbalance found in real-world systems, was split 60/30 into training and test sets. These were used for preprocessing, feature engineering, and training the Isolation Forest model. As shown in Figure 2 (b), the model successfully identified known anomalies within the test set, demonstrating its capability to detect abnormal behavior in a predominantly normal signal stream.
 
For \textbf{inference} validation, a 10-minute stress simulation was launched using \verb|cassandra-traffic|, applying a model previously trained on two weeks of normal telemetry. The inference module, running in event-driven mode, successfully detected injected anomalies, as evidenced by its internal logs and the anomaly entries recorded in InfluxDB. Figure 3 confirms this behavior, showing both log traces and stored anomalies accessible via the database’s Web UI.

\begin{figure}[htbp]
    \centering
    \includegraphics[width=0.77\linewidth]{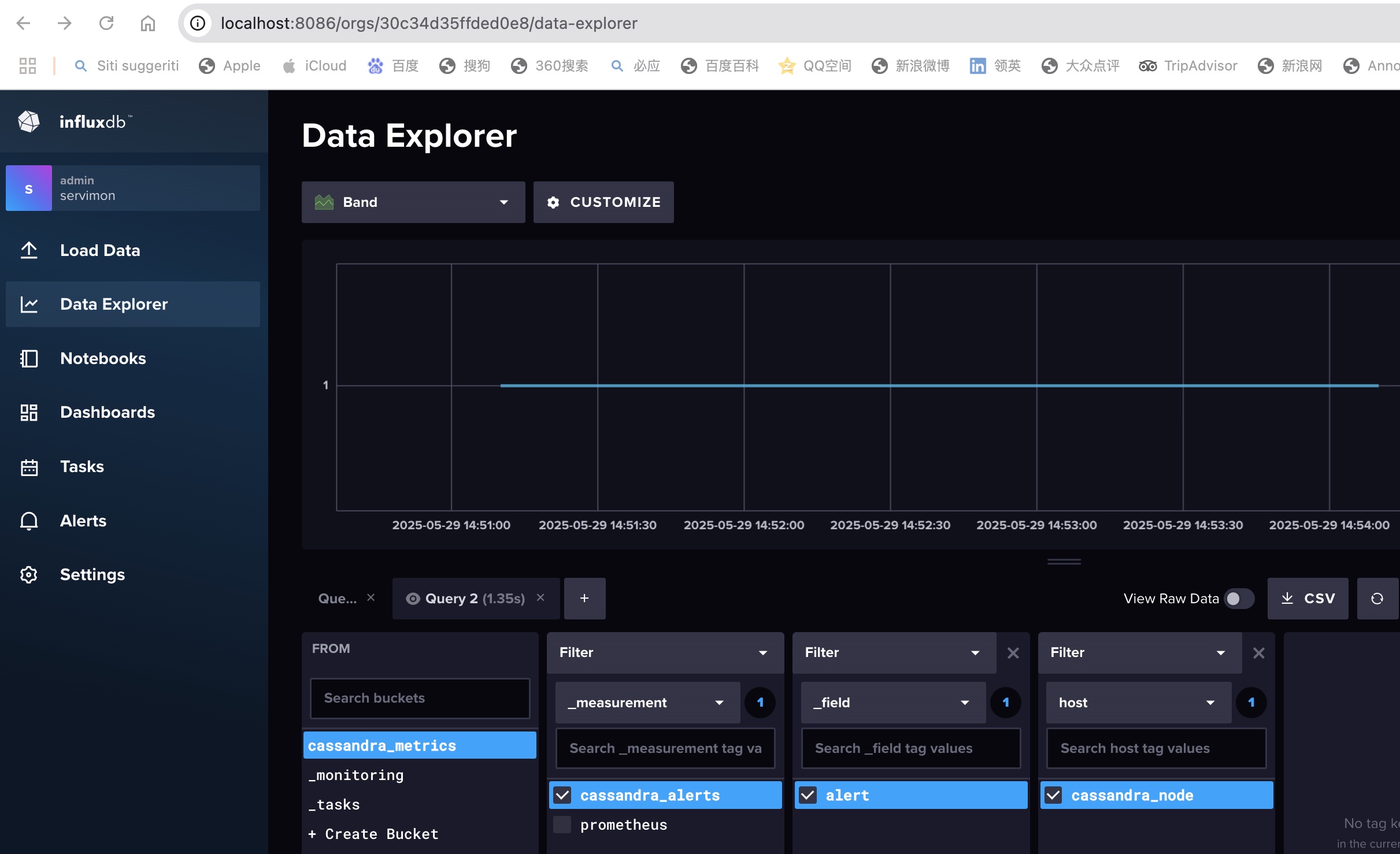}
    \caption{Anomalies shown in the browser}
    \label{fig:browser-alert}
\end{figure}

\begin{figure}[htbp]
    \ContinuedFloat
    \centering
    \includegraphics[width=0.77\linewidth]{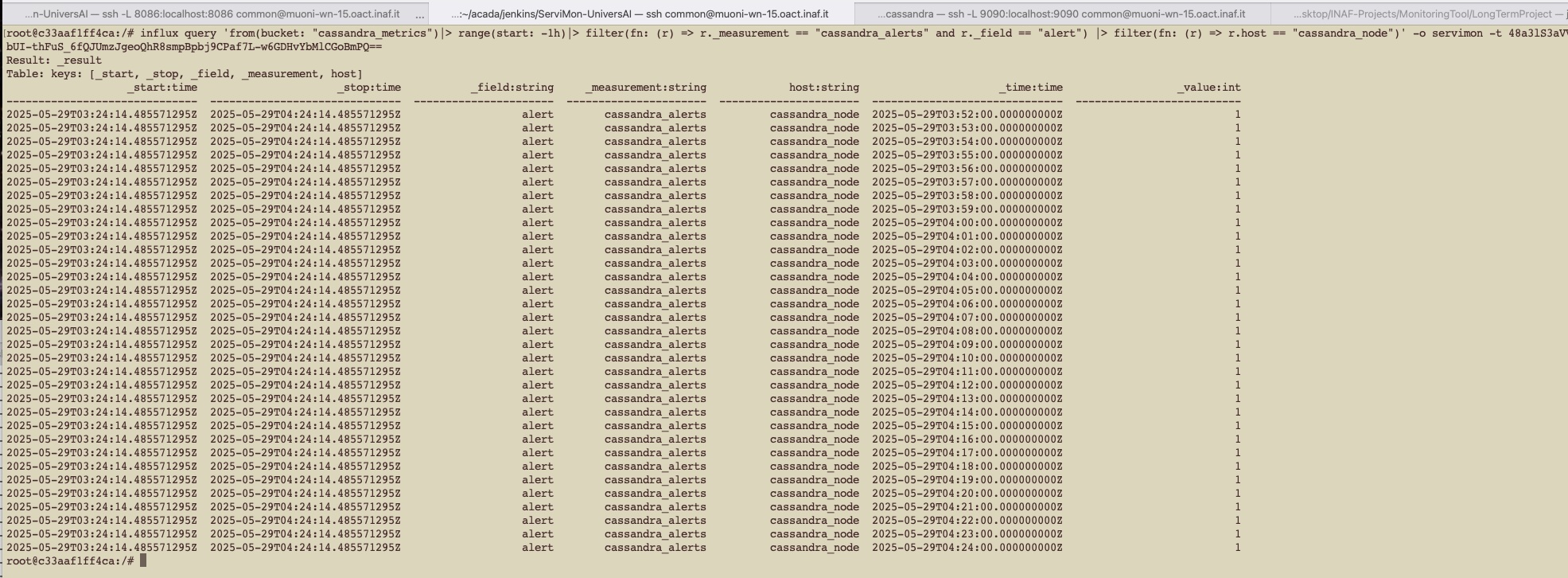}
    \caption{Anomaly detection on log files}
    \label{fig:table-alert}
\end{figure}

\section{Conclusion}

Originally developed to provide a monitoring system for astrophysical applications, ServiMon has been enhanced with an additional module focused on predictive maintenance.
This work has presented the basic architecture of ServiMon, detailing the machine learning module introduced and highlighting the overall system design as well as the main interactions among its components.
Preliminary tests and results demonstrate the validity of the proposed approach and the potential of the system.
Future developments include the integration of alternative algorithms to Isolation Forest—selected in this phase for its execution speed and ease of implementation—within the predictive maintenance module, as well as the use of real-world data to evaluate the system's functionality and performance.

\section{ACKNOWLEDGMENT}
This work is supported by ICSC – Centro Nazionale di Ricerca in High Performance Computing, Big Data and Quantum Computing, funded by the European Union – NextGenerationEU.


\begin{thebibliography}{99}

\bibitem{Giuliani}
Andrea Giuliani, 2023, arXiv, \url{https://arxiv.org/abs/2303.14079}

\bibitem{Actis2011}
Actis, M., Agnetta, G., Aharonian, F., et al., 2011, Experimental Astronomy, 32, 193, DOI 10.1007/s10686-011-9247-0

\bibitem{Munari2025}
K. Munari, A. Costa, F. Incardona, E. Mastriani, S. Spinello, et al.,
"Enhancing CTAO Monitoring and Alarm Subsystems in Distributed Environments Using ServiMon,"
in Proceedings of the 38th International Cosmic Ray Conference (ICRC 2025), 2025.

\bibitem{Docker}
Docker Inc., "Docker: Enterprise Container Platform", 2013. Available at: \url{https://www.docker.com}

\bibitem{RabensteinVolz}
Bjorn Rabenstein and Julius Volz, 2015, USENIX Association, \url{https://github.com/prometheus}

\bibitem{Grafana}
Grafana Labs, 2025, Grafana, \url{https://grafana.com}

\bibitem{GrafanaLoki}
Grafana Labs Loki, 2025, Grafana Loki, \url{https://grafana.com/oss/loki}

\bibitem{Promtail}
Grafana Labs Promtail, 2025, Promtail, \url{https://grafana.com/docs/loki/latest/clients/promtail/}

\bibitem{ApacheCassandra}
The Apache Software Foundation Cassandra, 2025, Apache Cassandra, \url{https://cassandra.apache.org/}

\bibitem{ApacheKafka}
The Apache Software Foundation Kafka, 2025, Apache Kafka, \url{https://kafka.apache.org/}

\bibitem{InfluxDB}
InfluxData, "InfluxDB: Open Source Time Series Database", 2013.\\
Available at: \url{https://www.influxdata.com}

\bibitem{IsolationForest}
F. T. Liu, K. M. Ting,  Z.-H. Zhou,
"Isolation Forest,"
in Proc. of the 8th IEEE International Conference on Data Mining (ICDM 2008),
Pisa, Italy, pp. 413–422, 2008. DOI: \url{10.1109/ICDM.2008.17}

\bibitem{Flux}
InfluxData, "Flux: A Lightweight Scripting Language for Querying Time Series Data", 2018.\\
Available at: \url{https://docs.influxdata.com/flux}

\bibitem{Telegraf}
InfluxData, "Telegraf: The Plugin-Driven Server Agent for Collecting \& Reporting Metrics", 2015.\\
Available at: \url{https://www.influxdata.com/time-series-platform/telegraf/}

\end{thebibliography}
\end{document}